\documentclass[aip,twocolumn,floatfix,reprint]{revtex4-1}

\draft 

\usepackage{graphicx}
\usepackage{dcolumn}
\usepackage{bm}
\usepackage{float}
\usepackage{color}
\usepackage{amsmath, amssymb}
\usepackage{mathtools}
\usepackage{textcomp}
\usepackage[caption = false]{subfig}

\begin{document}

\title{Glassy dynamics from generalized mode-coupling theory: existence and uniqueness of solutions for hierarchically coupled integro-differential equations}

\author{Rutger A.~Biezemans}
 \affiliation{%
	 Theory of Polymers and Soft Matter, Department of Applied Physics, Eindhoven University of Technology, 5600 MB Eindhoven, The Netherlands
}%
 \affiliation{%
	 Institute for Complex Molecular Systems, Eindhoven University of Technology, 5600 MB Eindhoven, The Netherlands
}%
\author{Simone Ciarella}
 \affiliation{%
	 Theory of Polymers and Soft Matter, Department of Applied Physics, Eindhoven University of Technology, 5600 MB Eindhoven, The Netherlands
}%
 \affiliation{%
	 Institute for Complex Molecular Systems, Eindhoven University of Technology, 5600 MB Eindhoven, The Netherlands
}%
\author{Onur \c{C}aylak}
 \affiliation{%
 Department of Mathematics and Computer Science, Eindhoven University of Technology, 5600 MB Eindhoven, The Netherlands
}%
 \affiliation{%
	 Institute for Complex Molecular Systems, Eindhoven University of Technology, 5600 MB Eindhoven, The Netherlands
}%
\author{Bj\"{o}rn Baumeier}
 \affiliation{%
 Department of Mathematics and Computer Science, Eindhoven University of Technology, 5600 MB Eindhoven, The Netherlands
}%
 \affiliation{%
	 Institute for Complex Molecular Systems, Eindhoven University of Technology, 5600 MB Eindhoven, The Netherlands
}%
\author{Liesbeth M.~C.~Janssen}
 \affiliation{%
	 Theory of Polymers and Soft Matter, Department of Applied Physics, Eindhoven University of Technology, 5600 MB Eindhoven, The Netherlands
}%
 \affiliation{%
	 Institute for Complex Molecular Systems, Eindhoven University of Technology, 5600 MB Eindhoven, The Netherlands
}%

\date{\today}

\begin{abstract}
Generalized mode-coupling theory (GMCT) is a  first-principles-based and systematically correctable framework to
predict the complex relaxation dynamics of glass-forming materials. The formal theory amounts to a hierarchy of infinitely many coupled integro-differential equations, which may be approximated using a suitable finite-order closure relation. Although previous studies have suggested that finite-order GMCT leads to well-defined solutions, and that the hierarchy converges as the closure level increases, no rigorous and general result in this direction is known. Here we unambiguously establish the existence and uniqueness of solutions to generic, schematic GMCT hierarchies that are closed at arbitrary finite order. We consider two types of commonly invoked closure approximations, namely mean-field and exponential closures. We also distinguish explicitly between overdamped and underdamped glassy dynamics, corresponding to hierarchies of first-order and second-order integro-differential equations, respectively. We find that truncated GMCT hierarchies closed under an exponential closure conform to previously developed mathematical theories, both in the overdamped and underdamped case, such that the existence of a unique solution can be readily inferred. Self-consistent mean-field closures, however, of which the well-known standard-MCT closure approximation is a special case, warrant additional arguments for mathematical rigour. We demonstrate that the existence of a priori bounds on the solution is sufficient to also prove that unique solutions exist for such self-consistent hierarchies. To complete our analysis, we present simple arguments to show that these a priori bounds must exist, motivated by the physical interpretation of the GMCT solutions as density correlation functions. Overall, our work contributes to the theoretical justification of GMCT for studies of the glass transition, placing this hierarchical framework on a firmer mathematical footing.
\end{abstract}

\maketitle 


\section{Introduction}
One of the major challenges in condensed matter physics is to understand the relaxation dynamics of glass-forming materials, such as supercooled liquids and dense colloidal suspensions \cite{Gotze1992RelaxationLiquids, Gotze2009ComplexTheory,debenedetti2001supercooled}. 
Arguably the most striking feature of glass formation is that the viscosity or relaxation time grows by many orders of magnitude upon mild variations in the temperature or density~\cite{Angell1995,debenedetti2001supercooled}; yet at the same time only weak changes in the material's structural properties are observed \cite{Royall2015,Janssen2018Mode-CouplingPrimerb}. It is this apparent disconnect between structural and dynamical properties that lies at the heart of the glass transition problem. Indeed, after decades of research, a direct, causal, and quantitative link between the structure and the dynamics of glass-forming materials is still lacking~\cite{Royall2015,Dyre2006,Cavagna2009,Tarjus2011,Berthier2011,binder2011glassy,Charbonneau2017}. 

Mode-coupling theory (MCT) is the only strictly first-principles, microscopically based theory that seeks to predict dynamic phenomena related to the glass transition using only structural information as input \cite{Kob2002,Gotze2009ComplexTheory, Janssen2014RelaxationTheory}. 
Briefly, MCT amounts to a time-dependent integro-differential equation for the so-called intermediate scattering function $F(\mathbf{k},t)$, i.e., a dynamic two-point correlation function which probes correlations in the density field for a given wavevector $\mathbf{k}$ and time $t$.
This equation is governed by a memory kernel that, to leading order, is written as an (a priori unknown) dynamic four-point density correlation function. MCT ad-hoc factorizes this four-point correlator into a product of two two-point density correlators $F(\mathbf{q},t)$ and $F(\mathbf{k-q},t)$, thereby yielding a closed set of coupled equations for the intermediate scattering functions at all possible wavevectors. Once the required structural information is given as MCT input -- in the simplest form only the static structure factor $S(\mathbf{k})\equiv F(\mathbf{k},0)$ -- the theory can then be solved self-consistently, thus effectively translating structural properties [$S(\mathbf{k})$] into dynamical [$F(\mathbf{k},t)$] ones \cite{Reichman2005Mode-couplingTheory}. 

Despite MCT's uncontrolled factorization approximation for the four-point correlators in the memory kernel, the theory has been remarkably successful in capturing important aspects of glassy dynamics. For example, MCT correctly predicts the emergence of a plateau in $F(\mathbf{k},t)$ upon supercooling, stretched exponential behavior, a time-temperature superposition principle, non-trivial scaling laws for the short-, intermediate- and long-time dynamics of $F(\mathbf{k},t)$, and complex reentrance phenomena \cite{Janssen2015MicroscopicPrinciples, Gotze2009ComplexTheory}. Quantitatively accurate predictions for $F(\mathbf{k},t)$ can also be obtained by means of a rescaling of temperature and density in the weakly to moderately supercooled regime \cite{Weysser2010StructuralSimulation}. However, due to the factorization approximation, MCT also suffers from several pathologies: the theory generally grossly overestimates the glass transition temperature~\cite{},  underestimates the violation of the Stokes-Einstein relation in the supercooled regime\cite{charbonneau2014hopping}, and fails to account for Arrhenius-type behavior and the concept of fragility~\cite{Ciarella2019UnderstandingPolymers}. Furthermore, MCT does not become exact in the mean-field limit of infinite spatial dimensions~\cite{Charbonneau13939,Schmid2010,PhysRevLett.104.255704}.

In order to remedy MCT's main uncontrolled approximation, i.e.\ the ad-hoc factorization of the four-point density correlators, Szamel proposed a new framework now referred to as generalized MCT (GMCT) \cite{Szamel2003ColloidalTheory}. Within GMCT, formally exact equations of motion for the (unknown) four-point correlators are derived, which yield a new integro-differential equation with a memory kernel that is dominated by six-point density correlators; these six-point correlators are subsequently governed by eight-point density correlation functions, and so on.
This hierarchical approach can, in principle, be continued up to arbitrary order. Previous work has shown that applying the factorization closure at the level of the six- and eight-point correlators, respectively, systematically improves the location at which the dynamical transition is predicted \cite{Szamel2003ColloidalTheory, Wu2005High-OrderTransition}. Continuing this procedure indefinitely extends GMCT to an infinite hierarchy of coupled integro-differential equations as obtained in Ref.~\onlinecite{Janssen2016GeneralizedOrder}.

Since, in general, no analytic solution to the infinite GMCT framework is known, a closure relation must be imposed at a finite level of the hierarchy so as to obtain a finite system that can be solved numerically. It is appealing to assume that such a hierarchy will converge with increasing closure level; however, there is no explicit small parameter in the theory which a priori warrants the neglect of higher order correlations, and hence it is not clear whether convergence can be generally achieved.
The authors of Ref.~\onlinecite{Janssen2016GeneralizedOrder} studied this question both numerically and analytically for so-called schematic GMCT frameworks, i.e.\ simplified versions of the full microscopic theory in which only one wavevector is included \cite{Bengtzelius1984DynamicsTransition, Leutheusser1984DynamicalTransition}. They considered two kinds of finite-order closure approximations, referred to as exponential and mean-field closures, respectively.
Remarkably, it was found that all studied GMCT hierarchies  converge uniformly with increasing closure level. For one specific schematic hierarchy with a known analytic infinite-order solution \cite{Mayer2006CooperativityTheory}, this uniform convergence could also be proven mathematically. Although these results are based on schematic GMCT models only, it is plausible that the fully microscopic framework shows similar convergence behavior; for standard MCT, it has already been firmly established that schematic models share many mathematical and physical properties with the wavevector-dependent theory \cite{Gotze2009ComplexTheory}. Indeed, all other reported numerical GMCT studies to date, either schematic or fully wavevector-dependent, have also suggested a systematic convergence of the GMCT hierarchy.


Here, we contribute to the theoretical justification of GMCT  by rigorously establishing the existence and uniqueness of solutions for generic, schematic GMCT hierarchies. Specifically, we will demonstrate that for both underdamped and overdamped dynamics, and for both exponential and mean-field closures applied at arbitrary order, a unique (time-dependent) solution exists. After briefly recapitulating the microscopic foundations of schematic GMCT in Sec.~\ref{GMCT}, we introduce the general integro-differential equations and closures relevant to schematic GMCT in Sec.~\ref{schematic}.
We will distinguish between the overdamped dynamics in Sec.~\ref{over-damped}  and the full underdamped dynamics in Sec.\ \ref{under-damped}. The equations for these different regimes have received different treatments in the mathematical literature. Previous studies have, however, not considered the applicability of these results to finite-order hierarchies as obtained in schematic GMCT under various closures. We show that exponential closures fit perfectly in previously developed studies, whereas mean-field closures require a new mathematical treatment for the case of underdamped dynamics. Therefore, in Sec.~\ref{under-damped}, we will introduce the modifications needed to complement previously published mathematical studies and we will construct complementary arguments to also establish existence and uniqueness of mean-field closure solutions. To this end, we assume the existence of a priori bounds on the density correlation functions, which will be motivated based on the physical background of GMCT. Finally, we summarize the main results and contributions of this paper in Sec.~\ref{conclusion}.

\section{Microscopic GMCT} \label{GMCT}

We first introduce the basic quantities and equations studied in microscopic (wavevector-dependent) MCT and GMCT; for more details, see e.g.\ Refs.~\onlinecite{Reichman2005Mode-couplingTheory, Janssen2015MicroscopicPrinciples}. For a system of $N$ particles with positions $\mathbf{r}_1(t),\dots,\mathbf{r}_N(t)$ at time $t$, the local density at a point $\mathbf{r}$ in space is given by
$$
\rho(\mathbf{r},t) = \sum\limits_{j=1}^N \delta(\mathbf{r}-\mathbf{r}_j(t)),
$$
where $\delta(\mathbf{r}-\mathbf{r}_j)$ denotes the delta function. The Fourier transform yields the density modes as a function of wavevector $\mathbf{k}$, 
$$
\rho(\mathbf{k},t) = \sum\limits_{j=1}^N e^{-i\mathbf{k}\mathbf{r}_j(t)}.
$$
The intermediate scattering function $F(\mathbf{k},t)$ is the time-dependent autocorrelation function of these density modes, 
\begin{equation}
F(\mathbf{k},t) = \frac{1}{N} \langle \rho(\mathbf{-k},0)\rho(\mathbf{k},t) \rangle, \label{Fdef}
\end{equation}
with the brackets denoting a canonical ensemble average. Using the definition of the static structure factor $S(\mathbf{k}) = \frac{1}{N} \langle \rho(\mathbf{k},0)\rho(\mathbf{-k},0) \rangle$, the dynamic density correlation functions can be normalized as 
\begin{equation}
\phi(\mathbf{k},t) = F(\mathbf{k},t)/S(\mathbf{k}). \label{phi_k_def}
\end{equation} 
The relaxation time of this function is a measure for the glassiness of the system, and hence it constitutes the key property predicted by standard MCT. In higher-order GMCT, the $2l$-point density correlation functions $\phi_l$ are also considered,
\begin{align}
    \phi_l&(\mathbf{k}_1,\dots,\mathbf{k}_l,t) = \nonumber \\
    &\frac{
    \langle \rho(-\mathbf{k}_1,0)\cdots \rho(-\mathbf{k}_l,0) \rho(\mathbf{k}_1,t) \cdots \rho(\mathbf{k}_l,t) \rangle
    }{
    \langle \rho(-\mathbf{k}_1,0)\cdots \rho(-\mathbf{k}_l,0) \rho(\mathbf{k}_1(0) \cdots \rho(\mathbf{k}_l,0) \rangle
    }. \label{phi_l_Def}
\end{align}
Note that these multi-point correlation functions probe density correlations over $l$ distinct wavevectors. In the GMCT framework of Refs.~\onlinecite{Szamel2003ColloidalTheory, Janssen2015MicroscopicPrinciples}, these satisfy the equations of motion
\begin{align}
    \ddot{\phi}_l&(\mathbf{k}_1,\dots,\mathbf{k}_l,t) + \nu \dot{\phi}_l(\mathbf{k}_1,\dots,\mathbf{k}_l,t) \nonumber \\
    &+ {\Omega_l}^2(\mathbf{k}_1,\dots,\mathbf{k}_l) \phi_l(\mathbf{k}_1,\dots,\mathbf{k}_l,t) 
    \nonumber \\
    &+ \int_0^t M_l(\mathbf{k}_1,\dots,\mathbf{k}_l,t-\tau) \dot{\phi}_l(\mathbf{k}_1,\dots,\mathbf{k}_l,\tau) d\tau = 0,
    \label{eq:eom}
\end{align}
where the dots denote time derivatives, $\nu$ represents a friction coefficient accounting for short-time dynamics, and the $\Omega_l$ are bare frequencies given by
\begin{equation}
    {\Omega_l}^2(\mathbf{k}_1,\dots,\mathbf{k}_l) = \frac{k_B T}{m} \left[ \frac{|\mathbf{k}_1|^2}{S(\mathbf{k}_1)} + \cdots + \frac{|\mathbf{k}_l|^2}{S(\mathbf{k}_l)} \right], \label{frequencies}
\end{equation}
with $k_B$ the Boltzmann constant, $T$ the temperature and $m$ the particle mass. The memory kernels $M_l$ in Eq.\ (\ref{eq:eom}) are given by
\begin{align}
    M_l&(\mathbf{k}_1,\dots,\mathbf{k}_l,t) = \frac{\rho k_B T}{16 m \pi^3} \sum\limits_{i=1}^l \frac{
    {\Omega_l}^2(\mathbf{k}_i)
    }{
    {\Omega_l}^2(\mathbf{k}_1,\dots,\mathbf{k}_l)
    } \nonumber \\
    &\times \int |\tilde{V}_{\mathbf{q,k_i-q}}|^2 S(\mathbf{q})S(\mathbf{k}_i-\mathbf{q})
    \label{memoryKernels}  \nonumber \\
    &\times \phi_{l+1}(\mathbf{q},\mathbf{k}_1 - \mathbf{q} \delta_{i,1}, \dots, \mathbf{k}_l - \mathbf{q}\delta_{i,l}, t) d\mathbf{q}, 
\end{align}
where $\rho$ is the bulk density, $\delta_{i,j}$ is the Kronecker delta, and the static vertices $\tilde{V}_{\mathbf{q,k_i-q}}$ are given by
\begin{equation}
\tilde{V}_{\mathbf{q,k_i-q}} = k_i^{-1} \left[
 (\mathbf{k_i} \cdot \mathbf{q}) c(q) + 
  \mathbf{k_i} \cdot (\mathbf{k_i-q}) c(|\mathbf{k_i-q}|)\right],
      \label{eq:vertices}
\end{equation}
with $k_i = |\mathbf{k}_i|$ and 
$c(q)=\rho^{-1}[1-1/S(q)]$ the
direct correlation function \cite{Hansen2013}.
Note in Eq.~(\ref{memoryKernels}) the explicit coupling of all wavevectors via the integral over $\mathbf{q}$, and the appearance of the $2(l+1)$-density correlator $\phi_{l+1}$ in the equation of motion for $\phi_{l}$. Hence, all dynamic multi-point density correlators are hierarchically coupled. The GMCT hierarchy of Eqs.\ (\ref{eq:eom})--(\ref{eq:vertices}) may subsequently be closed at arbitrary order, which will be discussed in more detail in Sec.\ \ref{schematic}; the closure approximation $\phi_2(\mathbf{k}_1,\mathbf{k}_2,t)\approx \phi_1(\mathbf{k}_1,t) \phi(\mathbf{k}_2,t)$ naturally recovers the standard MCT equations. 

As explained in Ref.\ \onlinecite{Janssen2015MicroscopicPrinciples}, the above microscopic GMCT equations are based on two remaining approximations: (i) so-called off-diagonal dynamic multi-point correlators are neglected, i.e.\ a set of $l$ distinct density modes at time $0$ is correlated only with \textit{the same set of wavevectors} at time $t$ [cf.\ Eq.\ (\ref{phi_l_Def})], and (ii) all \textit{static} multi-point density correlations, i.e.\ higher-order generalizations of the static structure factor, are factorized into products of $S(\mathbf{k})$. That is, all relevant microstructural information of the system is assumed to be contained in $S(\mathbf{k})$, but in principle one may also include higher-order structural correlators as additional theory input. Both of these approximations are implicitly also employed in standard MCT; importantly, the key improvement of GMCT is to avoid the factorization for the dynamic multi-point correlators $\phi_l$ in the memory kernel. The final GMCT equations of motion, Eq.\ (\ref{eq:eom}), are subject to the boundary conditions $\phi_l(\mathbf{k}_1, ... ,\mathbf{k}_l,t=0)=1$ and $\dot{\phi}_l(\mathbf{k}_1, ... ,\mathbf{k}_l,t=0)=0$ for all $l$.

\section{Schematic GMCT} \label{schematic}

Schematic mode-coupling theories reduce the full equations of motion to a simpler form by dropping all explicit wavevector dependence \cite{Bengtzelius1984DynamicsTransition, Leutheusser1984DynamicalTransition}.
As discussed by Bengtzelius \textit{et al}.\ \cite{Bengtzelius1984DynamicsTransition}, the memory kernel is dominated by the main peak of the static structure factor and, since the bifurcation point is the same in a scalar-valued theory, a schematic approach that neglects all other wavevectors is justified. 
A similar argument can also be made for higher-order GMCT; more details on this analysis will be presented in a separate publication.
Within schematic GMCT, the correlation functions $\phi_l(\mathbf{k}_1,\dots,\mathbf{k}_l,t)$ are approximated by $\psi_l$ in the following infinite hierarchy of underdamped coupled integro-differential equations \cite{Mayer2006CooperativityTheory,Janssen2016GeneralizedOrder}:
\begin{align} \label{GMCT2ndDeriv}
	\begin{cases}
		\ddot{\psi}_l(t) +  \zeta\dot{\psi}_l(t) + \mu_l \psi_l(t) \\ \quad + \lambda_l \int\limits_0^t \psi_{l+1}(t-\tau) \dot{\psi}_l (\tau) d\tau = 0, \\
		\psi_l(0) = 1, \, \dot{\psi}_l(0) = 0.
	\end{cases}
\end{align}
Mathematically, at each level $l \in \mathbb{N}$ in the hierarchy, $\psi_l(t)$ is a real-valued function defined on $[0,\infty)$. Further, $\zeta>0$ is an effective friction coefficient and $\mu_l$ represents a frequency for the schematic approximation to GMCT, which is positive in accordance with the frequencies of Eq.~(\ref{frequencies}). All $\lambda_l > 0$ represent the effective memory kernel weight in the equation at level $l$, effectuating a replacement of $M_l$ given in Eq.~(\ref{memoryKernels}) by $\lambda_l \psi_{l+1}$. In the overdamped limit, assuming the second-order time derivative can be neglected in comparison to the other terms and rescaling to $\zeta = 1$, the equations take the form
\begin{align} \label{GMCT1stDeriv}
	\begin{cases}
		\dot{\psi}_l(t) + \mu_l \psi_l(t) + \lambda_l \int\limits_0^t \psi_{l+1}(t-\tau) \dot{\psi}_l (\tau) d\tau = 0, \\
		\psi_l(0) = 1.
	\end{cases}
\end{align}

In Ref.~\onlinecite{Janssen2014RelaxationTheory}, the study of different choices of the parameters $\mu_l$ and $\lambda_l$ reveals that infinite-order schematic GMCT can predict avoided, discontinuous, and continuous glass transitions. Moreover, it was shown that the predictions of schematic standard MCT can also be reproduced by a full GMCT hierarchy by fitting the parameters to a certain plateau height and relaxation time.

In the absence of an analytic solution as $l\rightarrow\infty$, a closure approximation must be used to obtain a finite system of equations that can be solved numerically. A closure at level $L$ means that we presuppose a specific formula for $\psi_{L+1}$ in terms of $\psi_1,\dots,\psi_L$, which closes the systems of Eq.~(\ref{GMCT2ndDeriv}) or Eq.~(\ref{GMCT1stDeriv}) for the first $L$ unknowns. Previous work has established, both numerically for structural glass formers \cite{Janssen2015MicroscopicPrinciples,LuoJanssenarxiv} and analytically for specific schematic models \cite{Mayer2006CooperativityTheory,Janssen2014RelaxationTheory,Janssen2016GeneralizedOrder}, that the predictions for the density correlators $\psi_1$ manifestly converge for increasing closure levels.

We will consider from a mathematical point of view two closures which are now commonly used in the GMCT literature \cite{Mayer2006CooperativityTheory,Janssen2016GeneralizedOrder,Janssen2015MicroscopicPrinciples}. Firstly, the so-called exponential closure assumes that 
\begin{equation}
\psi_{L+1} \equiv 0.
\label{eq:expclosure}
\end{equation}
Note that this is essentially a simple truncation of the hierarchy.
With this closure, one then finds immediately for $\psi_L$ in Eq.\ (\ref{GMCT1stDeriv}) the explicit formula $\psi_L(t) = e^{-\mu_L t}$. Secondly, we will consider so-called mean-field closures of the form
\begin{equation}
\psi_{L+1} = \prod\limits_{l=1}^L (\psi_l)^{p_l}
\label{MFdef}
\end{equation}
where for $l=1,2,\dots,L$ the $p_l$ are chosen in $\mathbb{N}\cup\{0\}$ such that 
$$
\sum\limits_{l=1}^L l\cdot p_l = L+1.
$$
Note that this always leads to a self-consistent set of GMCT equations.
A typical example of a mean-field closure is 
\begin{equation}
    \psi_{L+1} = \psi_1 \psi_L. \label{MFexample}
\end{equation} 
When $L=1$, the only possibility $m_1=2$ corresponds to the standard-MCT-based $F_2$ model of Ref.~\onlinecite{Leutheusser1984DynamicalTransition}. Finally, we mention that earlier GMCT studies \cite{Janssen2015MicroscopicPrinciples,Janssen2016GeneralizedOrder} suggest that the exponential and mean-field closures at a given order constitute a lower and upper bound to the infinite-order solution, respectively.

We note that the hierarchies (\ref{GMCT2ndDeriv}) and (\ref{GMCT1stDeriv}), when closed at some level $L$, can be cast in the form of the equations of schematic standard MCT generalized to vector-valued functions. In this case the memory kernel $M_L$ at level $L$ is suitably defined in terms of  $\psi_1,\dots,\psi_{L}$ to translate the closure relation chosen. We shall introduce this notation when recalling the more abstract setting of Ref.~\onlinecite{Saal2013Well-posednessEquations} in Sec.~\ref{under-damped}. Since there is no structural difference in the integro-differential equations, mathematical results often generalize from scalar- to vector- or matrix-valued variants of the standard MCT equation \cite{Gotze1995GeneralEquations, Franosch2002CompletelyMixtures, Saal2013Well-posednessEquations}. 

To determine whether a unique solution exists for both types of GMCT closures at arbitrary (but finite) order, and for both overdamped and underdamped dynamics, we build upon earlier mathematical studies of coupled integro-differential equations. Since these previous works have used different techniques for first- and second-order integro-differential equations, the treatment of the over- and underdamped GMCT equations requires different conditions to be checked. Therefore, the existence and uniqueness questions for the systems of Eq.~(\ref{GMCT2ndDeriv}) and Eq.~(\ref{GMCT1stDeriv}) are discussed separately in the next sections. 

We note that previous work by Franosch has also obtained the existence of long-time limits for correlation functions described by the mode-coupling theory\cite{Franosch_2014}. These results might be generalized to GMCT. 
Finally, let us also note that existence and uniqueness for solutions to some classes of mode-coupling equations was obtained by Haussmann by means of the convergence of an iteration sequence, sharing similarities with the analysis that we mention in Sec.~\ref{over-damped}\cite{Haussmann1990SomeEquations}. The possible extension of this analysis to GMCT is not considered in the present paper.

\section{Overdamped dynamics} \label{over-damped}
We first turn our attention to an existence and uniqueness result of solutions for the overdamped system of Eq.~(\ref{GMCT1stDeriv}) with one of the closures at some level $L$ as described above. A closely related system was studied by Götze and Sjögren in Ref.~\onlinecite{Gotze1995GeneralEquations}, where the system is written in the form 

\begin{align} \label{GotzeSystem}
	\begin{cases}
		\dot{\psi}_n(t) = - \mu_n \psi_n(t) - \int\limits_0^t m_n(t-\tau) \dot{\psi}_n (\tau) d\tau, \\
		\psi_n(0) = 1.
	\end{cases}
\end{align}
Here, the kernel function $m_n$ is of the form $m_n(t) = G_n(\psi_1(t),\dots,\psi_L(t))$ for some functions $G_n$ describing the coupling and the closure chosen (we detail the case of Eq.~(\ref{MFexample}) below). By an iteration procedure for a linearized equation in the case $L=1$, studied via the Laplace transform, it is proved in Ref.~\onlinecite{Gotze1995GeneralEquations} that a unique solution exists in the class $C^1([0,T],\mathbb{R})$ for arbitrary final times $T>0$ under the condition that the corresponding one-dimensional kernel function $G_1$ is absolutely monotone on an interval $[0,1+\delta)$ for some $\delta>0$. This one-dimensional case corresponds to the explicit proof for schematic MCT, where indeed $L=1$ and $\psi_2 = {\psi_1}^2$ is the standard $F_2$ model studied in e.g.\ Ref.~\onlinecite{Leutheusser1984DynamicalTransition}. The same method of proof applies to each of the $L$ components of Eq.~(\ref{GotzeSystem}) that represents schematic GMCT. Reference \onlinecite{Gotze1995GeneralEquations} provides the existence of unique solutions $\psi_1,\dots,\psi_L$ in $C^1([0,T],\mathbb{R})$ for arbitrary times $T>0$ under the condition that $G_1,\dots,G_L$ all be absolutely monotone in each of their $L$ variables on $[0,1+\delta)^L$ for some $\delta>0$, i.e., for all $1 \leq l \leq L$,
\begin{equation}
\begin{split}
    \text{for all } j_1,&\dots,j_L \in \mathbb{N} \text{ and } x_1,\dots,x_L \in [0,1+\delta) : \\
    &\frac{\partial^{j_1}}{{\partial x_1}^{j_1}}\dots\frac{\partial^{j_L}}{{\partial x_L}^{j_L}} G_l(x_1,\dots,x_L) \geq 0.
\end{split} \label{GotzeCond}
\end{equation}

These conditions are clearly satisfied for both the exponential and mean-field closures of the GMCT hierarchy, since these closures yield monomial functions with positive coefficients for the $G_l$. For example, the mean-field closure of Eq.~(\ref{MFexample}) is described by $G_l(x_1,\dots,x_L) = \lambda_l x_{l+1}$ if $1\leq l \leq L-1$ and $G_L(x_1,\dots,x_L) = \lambda_L x_1 x_L$. All partial derivatives are positive if $x_1,\dots,x_L$ are positive, which yields the condition (\ref{GotzeCond}). Hence the existence and uniqueness question for overdamped GMCT at arbitrary order is fully covered by the theorem of Ref.~\onlinecite{Gotze1995GeneralEquations}.

Let us note that the exponential closure, included in the above analysis for completeness, can also be studied differently. Indeed, since Eq.~(\ref{eq:expclosure}) yields an explicit expression for $\psi_L$, the memory kernel of Eq.~(\ref{GotzeSystem}) is explicit at level $L-1$ and the theory of integral equations can be applied to obtain a solution $\psi_{L-1}$ that is sufficiently well behaved to make the same theory apply to the equation at level $L-2$. One iterates this reasoning and finally concludes by existence and uniqueness for $\psi_1$.

It is also known that the exponential closure leads via the Laplace transform to a truncated continued fraction representation for the Laplace transform of $\psi_1$, which can also be used to infer more properties of the solution. For studies that explicitly treat the Laplace transform, we refer to Refs.~\onlinecite{Haussmann1990SomeEquations, Gotze1995GeneralEquations, Franosch2002CompletelyMixtures}.

\section{Underdamped dynamics} \label{under-damped}
Let us now turn to the underdamped system of Eq.~(\ref{GMCT2ndDeriv}) closed at some level $L$. To simplify notation, the vector-valued analogue of Eq.~(\ref{GMCT2ndDeriv}) is introduced below in Eq.~(\ref{SaalEq2}). Since the structure of the integro-differential equation is unchanged, the mathematical analysis of the vector-valued variant of schematic MCT is indeed a direct generalization of the scalar case in Ref.~\onlinecite{Gotze1995GeneralEquations}. In Ref.~\onlinecite{Saal2013Well-posednessEquations}, only underdamped dynamics for the vector case are treated. We note that Franosch and Voigtmann extended the analyses of Ref.~\onlinecite{Gotze1995GeneralEquations} to matrix-valued MCT for the study of mixtures\cite{Franosch2002CompletelyMixtures}.

First we consider for $\Psi = (\psi_1,\dots,\psi_L)$ the integro-differential equation
\begin{align}
\begin{cases}
	\lambda \ddot{\Psi}(t) + \dot{\Psi}(t) + \Psi(t) \\ \quad + \int \limits_0^t m \left(\Psi(t-\tau) \right) \dot{\Psi}(\tau) d\tau = f(t),\\
    \Psi(0) = \Psi_0, \, \dot{\Psi}(0) = \Psi_1,
\end{cases} \label{SaalEq}
\end{align}
where $\Psi_0$ and $\Psi_1$ are arbitrary vectors in $\mathbb{R}^L$, $m \in C^1(\mathbb{R}^L,\mathbb{R}^{L \times L})$ and $f \in C([0,\infty),\mathbb{R}^L)$. 

The following local existence and uniqueness result is proved in Ref.~\onlinecite{Saal2013Well-posednessEquations}: there exists a (usually unknown) $T>0$ and a unique solution $\Psi$ of class $C^2$ on the interval $[0,T]$ satisfying Eq.~(\ref{SaalEq}). This local result is extended to global existence and uniqueness up to arbitrary $T>0$ under the following condition of linear growth:
\begin{equation}
\begin{split}
    \text{There exists } &c>0 \text{ such that } \forall x,y\in\mathbb{R}^L :\\
    &|m(x)y| \leq c(1+|x|)|y|.
\end{split} \label{linearCondition}
\end{equation}

We will now describe the tools used in the proof of Ref.~\onlinecite{Saal2013Well-posednessEquations}, in order to introduce the modifications needed for the application to the system of Eq.~(\ref{GMCT2ndDeriv}). Equation (\ref{SaalEq}) is written in the form  $(\mathcal{L}+\mathcal{N})\Psi = (f,\Psi_0,\Psi_1)$ with the linear operator $\mathcal{L} : \Psi \mapsto (\lambda \ddot{\Psi} + \dot{\Psi} + \Psi, \Psi(0), \dot{\Psi}(0))$ and the non-linear operator $\mathcal{N}$ defined by $(\mathcal{N}\Psi)(t) = \left(\int_0^t m \left(\Psi(t-\tau) \right) \dot{\Psi}(\tau) d\tau, 0, 0\right)$. The fact that $\mathcal{L}$ is a one-to-one mapping between $\Psi$ and the initial conditions is a well-known fact from the study of ordinary differential equations. Fredholm theory, the abstract functional analytic theory developed for the study of integral equations, is used in Ref.~\onlinecite{Saal2013Well-posednessEquations} to establish invertibility of the perturbed operator $\mathcal{L}+\mathcal{N}$.  The invertibility of $\mathcal{L}+\mathcal{N}$ for functions $\Psi$ defined on a time interval $[0,T]$ translates precisely to the existence of a unique solution to Eq.~(\ref{SaalEq}) on $[0,T]$ for arbitrary $f, \Psi_0$ and $\Psi_1$. The theory of Ref.~\onlinecite{Zeidler1990NonlinearApplications} allows to deduce the invertibility from estimates on $\mathcal{L}$ and $\mathcal{N}$, which is the approach followed in Ref.~\onlinecite{Saal2013Well-posednessEquations}.

In order to fully describe Eq.~(\ref{GMCT2ndDeriv}), all these results need to be extended to the system
\begin{align}
\begin{cases}
	\lambda \ddot{\Psi}(t) + \dot{\Psi}(t) + \omega \Psi(t)\\ \quad + \int \limits_0^t m \left(\Psi(t-\tau) \right) \dot{\Psi}(\tau) d\tau = f(t),\\
    \Psi(0) = \Psi_0, \, \dot{\Psi}(0) = \Psi_1,
\end{cases} \label{SaalEq2}
\end{align}
where $\omega$ is a constant diagonal $L\times L$ matrix with positive diagonal entries. Indeed, taking 
\begin{equation*}
    \omega = 
    \begin{pmatrix}
    \mu_1 & &0\\
    & \ddots & \\
    0 & &\mu_L
    \end{pmatrix}, \ 
    m(\Psi) = 
    \begin{pmatrix}
    \lambda_1 \psi_2 & &0\\
    & \ddots & \\
    0 & &\lambda_L \psi_{L+1}
    \end{pmatrix}
\end{equation*}
with $f\equiv 0$ and the initial conditions
$$
\Psi_0 = (1,\dots,1), \Psi_1 = (0,\dots,0)
$$
leads to the equations of schematic GMCT introduced in Sec.~\ref{schematic}, Eq.~(\ref{GMCT2ndDeriv}).

To establish existence and uniqueness for underdamped GMCT, we recognize that the above modifications amount to replacing the operator $\mathcal{L}$ by a new linear operator $\mathcal{L}_\omega$, $$\mathcal{L}_\omega \Psi = (\lambda \ddot{\Psi} + \dot{\Psi} + \omega\Psi, \Psi(0), \dot{\Psi}(0)).$$ Since all $\mu_l$ are positive, this introduces no difficulties in obtaining the same estimates as used in Ref.~\onlinecite{Saal2013Well-posednessEquations} to show invertibility of the operator $\mathcal{L}_\omega+\mathcal{N}$. For the exponential closures, existence of a unique global solution then follows directly, since $m$ can in this case be written as $m(x) = \operatorname{diag}(\lambda_1 x_2,\lambda_2 x_3,\dots,\lambda_{L-1}x_{L},0)$ and clearly satisfies the linearity condition (\ref{linearCondition}) by virtue of the Cauchy-Schwarz inequality. We note, however, that the exponential closure can again be treated iteratively following the comments at the end of Sec.~\ref{over-damped}.

In the case of any mean-field closure, however, condition (\ref{linearCondition}) is violated due to a product term in the assumption for $\psi_{L+1}$, e.g.\ $\psi_1\cdot\psi_L$ for the mean-field closure of Eq.~(\ref{MFexample}). Therefore we construct an additional argument to complement the mathematical literature with the needs of the physical setting of the mean-field closure. The argument relies on a bound on $\Psi$, which will be discussed in the next section for the specific context of GMCT. We can then cut off the kernel function $m$ to overcome the violation of condition (\ref{linearCondition}) by the following procedure.

To be precise in our argument, we first state the two hypotheses that we shall use on top of the existence and uniqueness results of Ref.~\onlinecite{Saal2013Well-posednessEquations}. We fix a mean-field closure relation and thereby the kernel function $m$ in Eq.~(\ref{SaalEq2}). We suppose that an a priori bound exist for solutions of Eq.~(\ref{SaalEq2}), Hypothesis (\ref{hyp1}):
\begin{align}
\label{hyp1}\tag{H1}
\begin{split}
    \text{There exists a constant }  B > 0  \text{ such } &\text{that,}\\
    \text{if } \Psi \text{ solves Eq.~(\ref{SaalEq2}) on } [0,T] ,\text{ we } &\text{have}\\ 
    |\Psi(t)| \leq B, \quad t \in [0,T].
\end{split}
\end{align}
$B$ is supposed to be independent of $T$. 

We note that such a bound is by no means obvious from a mathematical point of view, based solely on Eq.~(\ref{SaalEq2}) itself, but can be expected to exist for all physical situations. Intuitively, this is because all the correlation functions $\phi_l(\mathbf{k}_1,\dots,\mathbf{k}_l,t)$ measure correlations in the density fluctuations over time with respect to the initial configuration; moreover, they are normalized by virtue of Eq.~(\ref{phi_l_Def}). On more mathematical grounds, due to the Cauchy-Schwarz inequality, one has for any $l$:
\begin{align}
\begin{split} \label{eq:CS}
    &\langle \rho(-\mathbf{k}_1,0)\cdots \rho(-\mathbf{k}_l,0) \rho(\mathbf{k}_1,t) \cdots \rho(\mathbf{k}_l,t)\rangle^2
    \leq \\
    &\quad \langle \rho(-\mathbf{k}_1,0)\cdots \rho(-\mathbf{k}_l,0) \rho(\mathbf{k}_1,0) \cdots \rho(\mathbf{k}_l,0)\rangle \cdot \\
    &\quad \langle \rho(-\mathbf{k}_1,t)\cdots \rho(-\mathbf{k}_l,t) \rho(\mathbf{k}_1,t) \cdots \rho(\mathbf{k}_l,t)\rangle.
\end{split}
\end{align}
Since the propagator commutes with the Hamiltonian, it holds that
\begin{align}
\begin{split}
    &\rho(-\mathbf{k}_1,t)\cdots \rho(-\mathbf{k}_l,t) \rho(\mathbf{k}_1,t) \cdots \rho(\mathbf{k}_l,t)\rangle = \\
    &\quad \langle \rho(-\mathbf{k}_1,0)\cdots \rho(-\mathbf{k}_l,0) \rho(\mathbf{k}_1,0) \cdots \rho(\mathbf{k}_l,0)\rangle,
\end{split}
\end{align}
so Eq.~(\ref{eq:CS}) simplifies to 
\begin{align}
\begin{split}
    &|\langle \rho(-\mathbf{k}_1,0)\cdots \rho(-\mathbf{k}_l,0) \rho(\mathbf{k}_1,t) \cdots \rho(\mathbf{k}_l,t)\rangle|
    \leq \\
    &\quad |\langle \rho(-\mathbf{k}_1,0)\cdots \rho(-\mathbf{k}_l,0) \rho(\mathbf{k}_1,0) \cdots \rho(\mathbf{k}_l,0)\rangle|.
\end{split}
\end{align}
According to Eq.~(\ref{phi_l_Def}), this implies $| \phi_l(\mathbf{k}_1,\dots,\mathbf{k}_l,t)| \leq 1$. Now $\psi_1(t),\dots,\psi_L(t)$ are to be interpreted as the wavevector-independent approximations of the $\phi_l$ at the main peak of the static structure factor, so we extend these bounds to uniform bounds on the $\psi_l(t)$.

To circumvent the nonlinear character of mean-field closures, we introduce a smooth cut-off function $\chi : \mathbb{R}^L \rightarrow \mathbb{R}$, compactly supported with $0\leq\chi\leq1$, equal to 1 on $[-B,B]^L$ and identically zero outside the box $[-B-1,B+1]^L$. Our second hypothesis is that replacing the kernel $m$ by the cut-off kernel $\chi m$ does not affect the existence of a priori bounds:
\begin{align}
\label{hyp2}\tag{H2}
\begin{split}
    \text{If in } &\text{Eq.~(\ref{SaalEq2}), } m \text{ is replaced by } \chi m,\\
    \text{Hypothesis } &\text{(\ref{hyp1}) still holds for the new system}\\
    &\text{with the same constant } B.
\end{split}
\end{align}
The motivation for this hypothesis is that the GMCT does not `see' any changes in the kernel function $m(\psi_1,\dots,\psi_L)$ if it is cut off for values that are never reached by $\psi_1,\dots,\psi_L$. Mathematically, however, the Hypotheses (\ref{hyp1}) and (\ref{hyp2}) are not equivalent. 

For the existence proof under Hypotheses (\ref{hyp1}) and (\ref{hyp2}) we now start from a local solution $\Psi$ to Eq.\ (\ref{SaalEq2}) on $[0,T]$ for a $T>0$. The existence of such a $T$ and the existence and uniqueness of a corresponding solution $\Psi$ are guaranteed by the local result of Ref.~\onlinecite{Saal2013Well-posednessEquations}. Let $\chi$ be defined as above. Then the function $\chi m$ satisfies Condition (\ref{linearCondition}). For example, in the case of the closure of Eq.~(\ref{MFexample}) we have explicitly for any $x,y \in \mathbb{R}^L$:

\begin{equation}
    \chi(x)m(x)y = 
    \begin{pmatrix}
        \lambda_1 \chi(x) x_2 y_1\\
        \lambda_2 \chi(x) x_3 y_2\\
        \vdots\\
        \lambda_{L-1} \chi(x) x_L y_{L-1}\\
        \lambda_L\chi(x) x_1 x_L y_L
    \end{pmatrix} \label{mBoundExplicit}
\end{equation}
and the only non-linearity occurs with respect to $x$ in the last component. This nonlinear contribution can now be seen to be compliant with condition (\ref{linearCondition}), since we have
\begin{align}
    |\lambda_L \chi(x) x_1 x_L y_L| &\leq \lambda_L \cdot (B+1) |x_L| \cdot |y_L| \nonumber\\ 
    &\leq \lambda_L (B+1) |x| \cdot |y|, \label{mBoundComponent}
\end{align}
exploiting that $\chi(x)x_1 \leq 1\cdot (B+1)$ when $x_1\leq B+1$ and $\chi(x)x_1 \equiv 0$ when $x_1>B+1$.

As a result, there exists a unique global solution $\underline{\Psi}$ to the cut-off variant of Eq.~(\ref{SaalEq2}), that is,
\begin{align}
\begin{cases}
    \lambda \ddot{\underline{\Psi}}(t) + \dot{\underline{\Psi}}(t) + \omega \underline{\Psi}(t) \\
    \quad + \int \limits_0^t (\chi m) \left(\underline{\Psi}(t-\tau) \right) \dot{\underline{\Psi}}(\tau) d\tau = f(t), \\
    \underline{\Psi}(0) = (1,\dots,1), \underline{\dot{\Psi}}(0) = (0,\dots,0) 
\end{cases} \label{SaalEqCut}
\end{align}
However, by introducing the cut-off function $\chi$, we changed the equation that is solved and $\underline{\Psi}$ and $\Psi$ could be different functions on the time interval $[0,T]$ (where $\Psi$ is defined). To compare the two, we combine the a priori bound of Hypothesis (\ref{hyp1}) with the very definition of $\chi$ to see that $\chi(\Psi(t))m(\Psi(t))$ is equal to $m(\Psi(t))$ for $t$ in $[0,T]$. Hence $\Psi$ also solves Eq.~(\ref{SaalEqCut}) on $[0,T]$ and $\Psi$ coincides with $\underline{\Psi}$ on this interval by uniqueness of $\underline{\Psi}$. 

For global existence, it remains to be seen that $\underline{\Psi}$ is in fact a global solution to the original problem of Eq.~(\ref{SaalEq2}). This is where we need to invoke Hypothesis (\ref{hyp2}). Then we have $|\underline{\Psi}(t)|\leq B$ for all $t$. This implies that $\chi(\underline{\Psi}(t))m(\underline{\Psi}(t))$ reduces to $m(\underline{\Psi}(t))$ for all $t$ and the global solution $\underline{\Psi}$ to Eq.~(\ref{SaalEqCut}) is also seen to be a global solution to the schematic GMCT system of Eq.~(\ref{SaalEq2}).

This argument has to be generalized to arbitrary mean-field closures (as opposed to the single example of Eq.~(\ref{MFexample}) explicitly treated above). We observe that Eq.~(\ref{mBoundComponent}) can indeed be generalized to any mean-field closure in order to establish the condition (\ref{linearCondition}), using Hypothesis (\ref{hyp1}) to obtain bounds on the schematic correlators $\psi_l$ (which play the role of $x$ in Eq.~(\ref{mBoundComponent})). The remainder of the argument can be used unchanged to obtain global existence and uniqueness for Eq.~(\ref{GMCT2ndDeriv}) under the additional assumption of a priori bounds as expressed by Hypotheses (\ref{hyp1}) and (\ref{hyp2}).

Finally, this procedure of constructing a global solution by means of a cut-off kernel does not guarantee uniqueness of the global solution. To obtain uniqueness, we can drop the Hypotheses (\ref{hyp1}) and (\ref{hyp2}) by again invoking the results proved in Ref.~\onlinecite{Saal2013Well-posednessEquations}. There, uniqueness is proved for arbitrary locally Lipschitz continuous kernels $m$. This is true for any mean-field or exponential closure because these lead to continuously differentiable kernels $m$, which are in particular locally Lipschitz continuous.


\section{Conclusion} \label{conclusion}

This paper investigates whether unique mathematical solutions can exist for coupled integro-differential equations as encountered in generalized mode-coupling theory of the glass transition. We have considered GMCT hierarchies with both overdamped and underdamped dynamics, closed under either an exponential or mean-field closure approximation at finite order. In the case of overdamped dynamics, Eq.~(\ref{GMCT1stDeriv}),  we have seen that both the exponential closure and any type of mean-field closure lead to kernel functions that meet the hypotheses of the work in Ref.~\onlinecite{Gotze1995GeneralEquations} and we conclude that the existence and uniqueness theory is mathematically completely rigorous for these closures. This is a fundamental theoretical result for GMCT studies where the system of Eq.~(\ref{GMCT1stDeriv}) is solved numerically in the absence of analytical solutions \cite{Janssen2016GeneralizedOrder, Ciarella2019UnderstandingPolymers}. 

In the case of underdamped GMCT dynamics, Eq.~(\ref{GMCT2ndDeriv}), the same result is obtained for exponential closures by applying the work of Ref.~\onlinecite{Saal2013Well-posednessEquations}. For mean-field closures, however, the hypotheses provided in the theorem concerned are violated as a result of product terms in Eq.~(\ref{MFdef}). In this case, only a local existence result can be deduced, which does not necessarily hold for arbitrarily large final times. 

A consideration of the physical origins of the GMCT equations leads towards a possible remedy for this incompatibility with the mathematical literature. Starting from the definition of the density correlations of interest, we have derived a bound on the schematic density correlator $\phi_l(\mathbf{k},t)$. Inspired by this result, we propose an argument based on an a priori bound on solutions of the system of Eq.~(\ref{GMCT2ndDeriv}). Adapting the equations in a way that exploits this bound, allows us to verify the hypotheses of Ref.~\onlinecite{Saal2013Well-posednessEquations} and to obtain global existence and uniqueness of solutions also for mean-field closures. It has to be pointed out, however, that such an a priori bound is not evident from a strictly mathematical point of view, and complementary mathematical studies could aim to derive one based solely on Eq.~(\ref{GMCT2ndDeriv}) in order to render this argument completely rigorous.

Another very interesting property of GMCT mentioned in the Introduction is its potential convergence to an analytical solution at infinite level \cite{Janssen2016GeneralizedOrder}.
A general result in this direction is unknown to the authors (except for one special case \cite{Mayer2006CooperativityTheory} for which existence and uniform convergence as $L\to\infty$ were proved\cite{Janssen2016GeneralizedOrder}) and would be a valuable additional result for the rigorous justification of GMCT. We note that from a numerical perspective, the considerable effort required for the high dimensional solution could be partly minimized by using more efficient integration techniques such as an optimized wavevector grid~\cite{Caraglio2020}. 

To conclude, the rigorous results obtained above are important theoretical foundations for the study of the structure-dynamics link in glass-forming matter, for which GMCT at increasing order is a promising, systematically correctable, and fully first-principles-based theory. Our results are fundamental for present and future studies that investigate the GMCT solutions often numerically, when analytical results are not available, as a means to study e.g.\ fragility and tunability of liquid models in schematic GMCT \cite{Janssen2014RelaxationTheory,Janssen2016GeneralizedOrder,Ciarella2019UnderstandingPolymers}. Existence and uniqueness of solutions is then essential. 

\section*{Acknowledgements}

It is a pleasure to thank Chengjie Luo and Andr\'{e}s Montoya Castillo for valuable discussions.

\bibliographystyle{ieeetr}
\bibliography{ref.bib}

\begin{thebibliography}{10}

\bibitem{Gotze1992RelaxationLiquids}
W.~G{\"{o}}tze and L.~Sj{\"{o}}gren, ``{Relaxation processes in supercooled
  liquids},'' {\em Rep. Prog. Phys.}, vol.~55, 1992.

\bibitem{Gotze2009ComplexTheory}
W.~G{\"{o}}tze, {\em {Complex Dynamics of Glass-Forming Liquids: a
  Mode-Coupling Theory}}.
\newblock Oxford University Press, 2009.

\bibitem{debenedetti2001supercooled}
P.~G. Debenedetti and F.~H. Stillinger, ``{Supercooled liquids and the glass
  transition},'' {\em Nature}, vol.~410, pp.~259--267, 2001.

\bibitem{Angell1995}
C.~A. Angell, ``{Formation of Glasses from Liquids and Biopolymers},'' {\em
  Science}, vol.~267, pp.~1924--1935, 1995.

\bibitem{Royall2015}
C.~P. Royall and S.~R. Williams, ``{The role of local structure in dynamical
  arrest},'' {\em Phys. Rep.}, vol.~560, pp.~1--75, 2015.

\bibitem{Janssen2018Mode-CouplingPrimerb}
L.~M.~C. Janssen, ``{Mode-Coupling Theory of the Glass Transition: A Primer},''
  {\em Front. Phys.}, vol.~6, p.~97, 2018.

\bibitem{Dyre2006}
J.~C. Dyre, ``{Colloquium: The glass transition and elastic models of
  glass-forming liquids},'' {\em Rev. Mod. Phys.}, vol.~78, pp.~953--972, 2006.

\bibitem{Cavagna2009}
A.~Cavagna, ``{Supercooled liquids for pedestrians},'' {\em Phys. Rep.},
  vol.~476, pp.~51--124, 2009.

\bibitem{Tarjus2011}
G.~Tarjus, ``{An overview of the theories of the glass transition},'' in {\em
  Dynamical Heterogeneities in Glasses, Colloids, and Granular Media} ({L.
  Berthier, G. Biroli, J.-P. Bouchaud, L. Cipelletti} and W.~van Saarloos,
  eds.), ch.~2, pp.~39--67, Oxford University Press, 2011.

\bibitem{Berthier2011}
L.~Berthier and G.~Biroli, ``{Theoretical perspective on the glass transition
  and amorphous materials},'' {\em Rev. Mod. Phys.}, vol.~83, pp.~587--645,
  2011.

\bibitem{binder2011glassy}
K.~Binder and W.~Kob, {\em {Glassy materials and disordered solids: An
  introduction to their statistical mechanics}}.
\newblock World Scientific, 2011.

\bibitem{Charbonneau2017}
P.~Charbonneau, J.~Kurchan, G.~Parisi, P.~Urbani, and F.~Zamponi, ``{Glass and
  Jamming Transitions: From Exact Results to Finite-Dimensional
  Descriptions},'' {\em Annu. Rev. Condens. Matter Phys.}, vol.~8,
  pp.~265--288, mar 2017.

\bibitem{Kob2002}
W.~Kob, ``{Course 5: Supercooled Liquids, the glass transition, and computer
  simulations},'' in {\em Slow relaxations and nonequilibrium dynamics in
  condensed matter} ({J. Barrat, M. Feigelman, J. Kurchan} and J.~Dalibard,
  eds.), pp.~199--269, Springer Berlin Heidelberg, 2002.

\bibitem{Janssen2014RelaxationTheory}
L.~M.~C. Janssen, P.~Mayer, and D.~R. Reichman, ``{Relaxation patterns in
  supercooled liquids from generalized mode-coupling theory},'' {\em Phys. Rev.
  E}, vol.~90, p.~052306, 2014.

\bibitem{Reichman2005Mode-couplingTheory}
D.~R. Reichman and P.~Charbonneau, ``{Mode-coupling theory},'' {\em J. Stat.
  Mech. Theory Exp.}, pp.~267--289, 2005.

\bibitem{Janssen2015MicroscopicPrinciples}
L.~M.~C. Janssen and D.~R. Reichman, ``{Microscopic Dynamics of Supercooled
  Liquids from First Principles},'' {\em Phys. Rev. Lett.}, vol.~115,
  p.~205701, 2015.

\bibitem{Weysser2010StructuralSimulation}
F.~Weysser, A.~M. Puertas, M.~Fuchs, and {Th. Voigtmann}., ``{Structural
  relaxation of polydisperse hard spheres: Comparison of the mode-coupling
  theory to a Langevin dynamics simulation},'' {\em Phys. Rev. E}, vol.~82,
  2010.

\bibitem{charbonneau2014hopping}
P.~Charbonneau, Y.~Jin, G.~Parisi, and F.~Zamponi, ``{Hopping and the
  Stokes--Einstein relation breakdown in simple glass formers},'' {\em Proc.
  Natl. Acad. Sci. U.S.A.}, vol.~111, pp.~15025--15030, 2014.

\bibitem{Ciarella2019UnderstandingPolymers}
S.~Ciarella, R.~A. Biezemans, and L.~M.~C. Janssen, ``{Understanding,
  predicting, and tuning the fragility of vitrimeric polymers},'' {\em Proc.
  Natl. Acad. Sci. U.S.A.}, vol.~116, 2019.

\bibitem{Charbonneau13939}
P.~Charbonneau, A.~Ikeda, G.~Parisi, and F.~Zamponi, ``{Dimensional study of
  the caging order parameter at the glass transition},'' {\em Proc. Natl. Acad.
  Sci. U.S.A.}, vol.~109, pp.~13939--13943, 2012.

\bibitem{Schmid2010}
B.~Schmid and R.~Schilling, ``{Glass transition of hard spheres in high
  dimensions},'' {\em Phys. Rev. E}, vol.~81, p.~41502, apr 2010.

\bibitem{PhysRevLett.104.255704}
A.~Ikeda and K.~Miyazaki, ``{Mode-Coupling Theory as a Mean-Field Description
  of the Glass Transition},'' {\em Phys. Rev. Lett.}, vol.~104, p.~255704, jun
  2010.

\bibitem{Szamel2003ColloidalTheory}
G.~Szamel, ``{Colloidal Glass Transition: Beyond Mode-Coupling Theory},'' {\em
  Phys. Rev. Lett.}, vol.~90, 2003.

\bibitem{Wu2005High-OrderTransition}
J.~Wu and J.~Cao, ``{High-Order Mode-Coupling Theory for the Colloidal Glass
  Transition},'' {\em Phys. Rev. Lett.}, vol.~95, 2005.

\bibitem{Janssen2016GeneralizedOrder}
L.~M.~C. Janssen, P.~Mayer, and D.~R. Reichman, ``{Generalized mode-coupling
  theory of the glass transition: Schematic results at finite and infinite
  order},'' {\em J. Stat. Mech. Theory Exp.}, vol.~2016, 2016.

\bibitem{Bengtzelius1984DynamicsTransition}
U.~Bengtzelius, W.~G{\"{o}}tze, and A.~Sj{\"{o}}lander, ``{Dynamics of
  supercooled liquids and the glass transition},'' {\em J. Phys. Condens.
  Matter}, vol.~17, pp.~5915--5934, 1984.

\bibitem{Leutheusser1984DynamicalTransition}
E.~Leutheusser, ``{Dynamical model of the liquid-glass transition},'' {\em
  Phys. Rev. A}, vol.~29, pp.~2765--2773, 1984.

\bibitem{Mayer2006CooperativityTheory}
P.~Mayer, K.~Miyazaki, and D.~R. Reichman, ``{Cooperativity beyond caging:
  Generalized mode-coupling theory},'' {\em Phys. Rev. Lett.}, vol.~97,
  pp.~1--4, 2006.

\bibitem{Hansen2013}
J.-P. Hansen and I.~R. McDonald, {\em {Theory of simple liquids}}.
\newblock Elsevier, 2013.

\bibitem{LuoJanssenarxiv}
C.~Luo and L.~M.~C. Janssen, ``{Generalized mode-coupling theory of the glass
  transition. I. Numerical results for Percus-Yevick hard spheres},'' {\em
  arXiv:1909.00428}, 2019.

\bibitem{Saal2013Well-posednessEquations}
M.~Saal, ``{Well-posedness and asymptotics of some nonlinear
  integro-differential equations},'' {\em J. Integral Equat. Appl.}, vol.~25,
  pp.~103--141, 2013.

\bibitem{Gotze1995GeneralEquations}
W.~G{\"{o}}tze and L.~Sj{\"{o}}gren, ``{General Properties of Certain
  Non-linear Integro-Differential Equations},'' {\em J. Math. Anal. Appl.},
  vol.~195, pp.~230--250, 1995.

\bibitem{Franosch2002CompletelyMixtures}
T.~Franosch and T.~Voigtmann, ``{Completely monotone solutions of the
  mode-coupling theory for mixtures},'' {\em J. Stat. Phys.}, vol.~109,
  pp.~237--259, 2002.

\bibitem{Franosch_2014}
T.~Franosch, ``Long-time limit of correlation functions,'' {\em Journal of
  Physics A: Mathematical and Theoretical}, vol.~47, p.~325004, jul 2014.

\bibitem{Haussmann1990SomeEquations}
R.~Haussmann, ``{Some properties of mode coupling equations},'' {\em Z.Phys.
  B}, vol.~79, pp.~143--148, 1990.

\bibitem{Zeidler1990NonlinearApplications}
E.~Zeidler, {\em {Nonlinear Functional Analysis and its Applications}},
  vol.~II/B: Nonlinear Monotone Operators.
\newblock Springer Science + Business Media, 1990.

\bibitem{Caraglio2020}
M.~Caraglio, L.~Schrack, G.~Jung, and T.~Franosch, ``{An improved integration
  scheme for Mode-coupling-theory equations},'' {\em arXiv:2007.07621}, 2020.

\end{thebibliography}


\end{document}